# Kitten or Panda? Measuring the Specificity of Threat Group Behaviors in Public CTI Knowledge Bases

Aakanksha Saha
TU Wien
Vienna, Austria
aakanksha.saha@seclab.wien

Martina Lindorfer
TU Wien
Vienna, Austria
martina@seclab.wien

Juan Caballero
IMDEA Software Institute
Madrid, Spain
juan.caballero@imdea.org

## Abstract

In recent years, the cyber threat intelligence (CTI) community has invested significant effort in building knowledge bases that catalog threat groups. These knowledge bases associate each threat group with its observed behaviors, including their Tactics, Techniques, and Procedures (TTPs) as well as the malware and tools they employ during attacks. However, the distinctiveness and completeness of such behavioral profiles remain largely unexplored, despite being critical for tasks such as threat group attribution. In this work, we systematically analyze threat group profiles built from two public CTI knowledge bases: MITRE ATT&CK and Malpedia. We first investigate what fraction of threat groups have group-specific behaviors, i.e., behaviors used exclusively by a single group. We find that only 34% of threat groups in ATT&CK have group-specific techniques, limiting the use of techniques as reliable behavioral signatures to identify the threat group behind an attack. The software used by a threat group proves to be more distinctive, with 73% of ATT&CK groups using group-specific software. However, this percentage drops to 24% in the broader Malpedia dataset. Next, we evaluate how group profiles improve when data from both sources are combined. While coverage improves modestly, the proportion of groups with group-specific behaviors remains under 30%. We then enhance profiles by adding exploited vulnerabilities and additional techniques extracted from threat reports. Despite the additional information, 64% of groups still lack any group-specific behavior. Our findings raise concerns about the specificity of existing behavioral profiles and highlight the need for caution, as well as further improvement, when using them for threat group attribution.

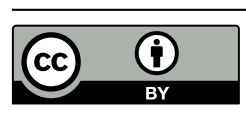

## 1 Introduction

In recent years, the cyber threat intelligence (CTI) community has explored the behavioral characteristics of threat groups (or threat actors), such as the Tactics, Techniques, and Procedures (TTPs) used to gain access, move laterally, maintain persistence, and exfiltrate data [2, 5, 20, 45, 56, 64]. Behavioral characteristics are thought to be distinctive, remain stable over time, and be able to link seemingly unrelated attacks from the same threat group [23]. In fact, when sufficiently distinctive, behavioral characteristics such as TTPs can serve as behavioral signatures that enable reliable attribution [39]. Models like the Pyramid of Pain [4] place TTPs at the top of the pyramid in terms of cost for an adversary to change them. Apart from TTPs, other behavioral characteristics also exist, for example, the software tools used by a threat group may be distinctive, particularly if the malware is developed in-house. Similarly, exploited vulnerabilities can be characteristic, especially when a group targets uncommon software or uses custom-developed exploits. Even the textual content used in campaigns can be characteristic of a threat group, with recent related work leveraging phishing SMS contents [34], ransomware notes [60], and cross-file-type features [49] to identify attacks from the same campaign and threat group.

We refer to a threat group's observed behaviors as its *group profile*. Accurate threat group profiles are fundamental to incident correlation, attribution, the development of behavioral detection rules, and proactive threat hunting. Group profiles can be constructed from threat reports, which typically describe, in natural language, analyses of specific attacks and their attribution to specific threat groups. Threat reports may come from a single source (e.g., a specific cybersecurity vendor) or be aggregated by CTI knowledge bases [40, 58] and sharing platforms [6, 20].

While the potential of such behavioral profiles has been widely acknowledged, few studies have examined how distinctive these profiles truly are or how complete our understanding of these behaviors is, particularly given the varying quality and scope of the data sources used to build them. One concern is that many behaviors in these profiles can be *generic*, i.e., used by many threat groups, thus providing little information about the groups using them. These include common techniques (e.g., spearphishing, malware autostart through registry keys), widely available software (e.g., abused penetration testing tools, open-source projects, malware kits sold on underground forums), and prevalent vulnerabilities (e.g., those affecting popular software with public exploits). Threat groups may acquire such tools and exploits for convenience, reduced operational costs, or to obscure attribution.

Recent work that interviewed threat analysts to gain insights into the attribution of advanced persistent threats (APTs) [50] and ransomware groups [61] showed that, although TTPs are used in the attribution process, they are often considered generic and inconsistent. Our work uses a data-driven approach to identify *group-specific* behaviors, i.e., behaviors used by a single threat group and therefore suitable for reliable attribution. Group-specific behaviors can consist of single behaviors or combinations of behaviors (e.g., pairs or triplets). When observed on a protected system, such group-specific behaviors serve as behavioral signatures that uniquely identify the responsible threat group. However, what fraction of threat groups exhibit group-specific behaviors remains an open question. Determining whether a behavior is group-specific requires not only analyzing the group that exhibits it, but also having comprehensive coverage of behaviors across other threat groups. Without this broader context, a behavior might appear group-specific when it is not. For example, as more threat reports become available, a behavior initially believed to be exclusive to group A may also be observed in group B, indicating it is not group-specific.

We investigate how distinctive threat group behavioral profiles in public knowledge bases are, how they can be expanded, and to what extent they support reliable threat group attribution. We focus on knowledge bases that are public (i.e., non-commercial), index many threat reports, are periodically updated, provide a taxonomy of threat groups, and organize information into group profiles. These profiles include basic group metadata (e.g., name, aliases, country), references to related threat reports, and descriptions of group behaviors mentioned in those threat reports. We find two knowledge bases satisfying those properties: *MITRE's Adversarial Tactics, Techniques, and Common Knowledge (ATT&CK)* [28] and *Malpedia* [15]. We exclude other projects (e.g., MISP threat actor galaxy [27], ThreatMiner [59], ORKL [38], and APTnotes [3]), as they collect threat reports and associate them with threat groups but do not extract or include behavioral information from those reports into their group profiles. Moreover, Malpedia incorporates data from the MISP threat actor galaxy, thereby indirectly covering it. We also exclude commercial services that provide specialized threat reports to paying customers [5], as their knowledge bases are proprietary. Both ATT&CK and Malpedia provide their own taxonomies of threat groups and associated software tools. A key distinction is that ATT&CK also includes a taxonomy of TTPs, integrating attack techniques directly into the group profiles. Based on ATT&CK and Malpedia, we answer the following questions:

**RQ1: What fraction of the threat groups in ATT&CK and Malpedia have group-specific behaviors?** We separately analyze the group profiles created using only the information available in ATT&CK and Malpedia. Identifying groups based on TTPs is challenging as only 52 (34.2%) groups in ATT&CK have group-specific techniques. It is somewhat easier to identify groups through the software they use, with 111 (73.0%) groups in ATT&CK having group-specific software. However, this percentage is significantly lower in Malpedia, where only 192 (24.0%) have group-specific software. This discrepancy stems from threat reports that disproportionately focus on a subset of high-profile groups, leaving less-known groups with sparse reporting to build their profiles. Combining techniques and software into joint profiles increases the groups with group-specific behaviors from 111 (73.0%) to 124 (81.6%).

**RQ2: How complementary is the information in both datasets? How much do group profiles improve when combining data from both sources?** Both datasets differ substantially in volume. Malpedia provides broader coverage of the threat landscape than ATT&CK, comprising 5.2 times more threat groups (800 vs. 152) and 4.2 times more software (3,367 vs. 794). This expanded scope stems from Malpedia indexing 16.9 times more threat reports (15,699 report URLs vs. 930). To assess their overlap, we normalize group and software names across the datasets. Both datasets have little overlap, with only 145 groups and 498 software entries in common. The corresponding Jaccard Index values are 17.7% for groups, 13.5% for software, and just 3.2% for report URLs. The low intersection indicates that each dataset captures a different view of the threat group landscape, highlighting their complementary nature.

We create joint group profiles using the data from both datasets, identifying 236 (29.2%) groups with group-specific behaviors, compared to 124 groups using only ATT&CK and 192 using only Malpedia. Despite combining the two datasets, more than 70% of groups exhibit no group-specific behavior.

**RQ3: What additional information currently not in ATT&CK and Malpedia could make threat group profiles more complete?** We examine how group profiles can be improved with additional information extracted from threat reports. First, we build vulnerability profiles for each threat group with the Common Vulnerabilities and Exposures (CVE) identifiers [44] of the vulnerabilities that the group has exploited. The number of groups with at least one group-specific vulnerability is 48 (31.6%) in ATT&CK, 112 (14.0%) in Malpedia, and 119 (14.7%) when combining both datasets. Thus, exploited vulnerabilities tend to be less distinctive than the software used but more distinctive than the techniques employed. Next, we extend the group profiles with additional technique identifiers extracted from the threat reports. Since Malpedia does not provide TTPs, this step allows us to extend its group profiles with techniques. Incorporating these extracted techniques increases the number of groups with group-specific behaviors from 52 (6.4%) to 68 (8.4%). Finally, we combine all available behavioral indicators, including techniques, extracted techniques from reports, software, and vulnerabilities, into unified group profiles, identifying 291 (36.0%) groups with at least one group-specific behavior. Despite leveraging all available information, a majority of groups (64%) have no group-specific behaviors.

To better understand the limitations of current group profiles, we also discuss the impact of under-reporting, i.e., incomplete coverage of threat group behaviors. We observe that the number of technique identifiers extracted from the ATT&CK threat reports is larger than the number of techniques officially cataloged in ATT&CK from those same reports. This discrepancy likely arises from the manual nature of the report analysis process by ATT&CK contributors, emphasizing the need for automated approaches to extract TTPs from threat reports [1, 17, 43]. We also observe that only 46.3% of techniques and 64.1% of software entries in ATT&CK, and just 28.6% of software in Malpedia, are currently associated with at least one threat group. The remaining entries were likely added to the taxonomies because they were observed being used by adversaries in the wild. However, their lack of association with specific threat groups highlights the incomplete coverage of group profiles.

**Artifacts.** Our code and data are available at: https://github.com/SecPriv/ThreatGroupCTI.

## 2 Dataset Comparison

For our analysis, we need public knowledge bases that index many threat reports, are periodically updated, provide a taxonomy of threat groups, and organize information into group profiles. While there exist many repositories of threat reports (e.g., ThreatMiner [59], APTnotes [3]) and reports can also be collected directly from security vendor blogs [7], we identified only three datasets that explicitly index threat reports by the threat groups they reference, making them suitable for building group profiles. Those three sources are MITRE's Adversarial Tactics, Techniques, and Common Knowledge (ATT&CK) [28], Malpedia [15], and the Malware Information Sharing Platform (MISP) threat actor galaxy [27]. Furthermore, because the MISP threat actor taxonomy is used as input to Malpedia, the two sources largely overlap, as shown later in this section. We focus on ATT&CK and Malpedia as our primary datasets.

In this section, we first detail the information in ATT&CK [28] and Malpedia [15], and then set the base for answering RQ2 by analyzing the extent of data overlap between the two datasets and assessing how their contents complement each other.

### 2.1 Datasets

**MITRE ATT&CK.** ATT&CK provides taxonomies of offensive and defensive techniques, software tools used by adversaries, and threat groups. The techniques taxonomy comprises three *domains*: Enterprise, Mobile, and Industrial Control Systems (ICS). Each domain defines a set of *tactics* that correspond to different steps in the kill chain, such as *Reconnaissance* (TA0043), *Persistence* (TA0003), and *Lateral Movement* (TA0008). Each tactic includes a set of *techniques*. For example, *Active Scanning* (T1595) and *Phishing for Information* (T1598) are techniques under the *Reconnaissance* tactic. Techniques can also contain *sub-techniques*. For example, T1204.002 corresponds to the *Malicious File* sub-technique under the *User Execution* (T1204) technique.

The threat group taxonomy covers nation-state actors, advanced persistent threats (APTs), and some large for-profit actors such as ransomware groups. A threat group profile includes a unique identifier, a name, a list of aliases (called *Associated Groups*), and the techniques and software used by the threat group.

Entries in the software taxonomy are categorized into *Tools* and *Malware*. Tools include commercial software (e.g., Cobalt Strike), open-source frameworks (e.g., Metasploit, Mimikatz), and built-in operating system (OS) tools (e.g., PsExec, ipconfig). Malware includes families specific to a single threat group (e.g., Carbanak) as well as malware kits available in underground markets and used by multiple threat groups (e.g., PoisonIvy RAT). The focus is on software used by APTs listed in the group taxonomy; however, ATT&CK also catalogs non-APT malware such as the Conficker worm [31] and the SimBad Android malware [32]. Each taxonomy entry contains URLs to threat reports related to the entry, such as reports describing the techniques and software used.

Since its first public release in 2015, MITRE has published a new ATT&CK version approximately every six months; the latest version at the beginning of our study, v15.1, was released in April 2024. Each version may add new taxonomy entries (e.g., groups, techniques, software), remove revoked entries, or mark entries as deprecated (i.e., to be revoked soon).

**Table 1: ATT&CK and Malpedia summary. Malpedia does not have a techniques taxonomy. The low intersection and Jaccard Index show that both datasets have little overlap. We use the union of both datasets to build group profiles.**

| Data | ATT&CK | Malpedia | ∩ | ∪ | Jaccard |
|---|---|---|---|---|---|
| Groups | 152 | 800 | 145 | 807 | 17.7% |
| Techniques | 839 | - | - | 839 | - |
| Software | 794 | 3,367 | 498 | 3,663 | 13.5% |
| Report URLs | 930 | 15,699 | 522 | 16,107 | 3.2% |
| Report FQDNs | 218 | 2,002 | 194 | 2,026 | 9.6% |
| Reports | 920 | 14,983 | 80 | 15,816 | 0.5% |

**Malpedia.** Malpedia provides taxonomies of threat groups and software. It does not provide a taxonomy of techniques nor reference the techniques in ATT&CK. Similar to ATT&CK, the threat group taxonomy focuses on APTs and nation-state actors. In contrast, the software taxonomy aims to cover any malware family, regardless of whether it is used by APTs or other types of attackers (e.g., for-profit actors). The software taxonomy also includes a few security tools (e.g., Cobalt Strike) but does not differentiate between malware and tools. Each taxonomy entry for a threat group or software includes URLs of threat reports related to the entry. We collect information about groups and software using the Malpedia API and metadata on threat reports (e.g., URL, title, author, publication date) from the provided BibTex file. Malpedia is updated daily by adding new bibliographic references labeled with the associated threat groups and software. We obtained Malpedia data on February 18, 2025.

**Malware Information Sharing Platform (MISP).** We also examined the MISP threat-actor galaxy but found substantial overlap with Malpedia, as Malpedia uses it as an input. In particular, we compared the threat report URLs in both datasets on August 6, 2025. On that date, the MISP galaxy contained 857 threat actors (the same as Malpedia) and 2,660 associated threat-report URLs. Of these, 91.39% of URLs and 95.99% of fully-qualified domain names (FQDNs) in those URLs are also present in Malpedia, indicating a high degree of overlap. Given this redundancy, we exclude MISP in our analysis, as Malpedia already provides a representative coverage of its content.

**Dataset comparison.** Table 1 summarizes the contents of both datasets. ATT&CK v15.1 contains 152 groups, 358 techniques, 481 sub-techniques, 794 software entries, and 930 URLs of threat reports from which those associations are extracted. Of the 358 techniques, 121 (33.8%) have at least one sub-technique, while 237 (66.2%) do not have sub-techniques. Among the total 839 techniques and sub-techniques, 637 (75.9%) belong to the Enterprise domain (202 techniques and 435 sub-techniques), 119 (14.2%) to Mobile (73 techniques and 46 sub-techniques), and 83 (9.9%) to ICS (83 techniques and no sub-techniques). For simplicity, in the remainder of this paper, we use the term *techniques* to refer to the combined set of 839 techniques and sub-techniques

In contrast, Malpedia does not include techniques; however, it is much larger, containing 800 groups (5.2x), 3,367 software (4.2x), and 15,699 (16.9x) report URLs. The report URLs in ATT&CK come from 218 domains, compared to 2,002 (9.2x) in the Malpedia URLs, indicating that Malpedia draws from a significantly more diverse set of sources (e.g., cybersecurity vendors and analyst blogs). We download the content of each URL, filter errors, and identify reports by the SHA256 of the downloaded content (most often an HTML page or a PDF document). In total, we downloaded 920 unique reports from the 930 ATT&CK URLs and 14,983 unique reports from 15,699 Malpedia URLs. We use the downloaded reports to extract additional information for extending the group profiles discussed in Section 4. Note that we perform deduplication at the report level (e.g., SHA-256 checksums). Because we build group profiles from the set of techniques or software extracted across all associated reports, near-duplicate reports describing the same campaign do not inflate the prominence of any particular behavior.

## 2.2 Dataset Intersection and Union

Here, we examine the datasets' overlap and the benefits of combining them. A key challenge is that group and software names differ across the datasets, so we created a mapping to align them.

Each knowledge base provides a name and a list of aliases for each threat group. We first normalized all names and aliases by converting them to lowercase, removing common prefixes and suffixes such as "group" or "framework," replacing "team" with a space, and converting "threat group" to "TG". Then, we compute the intersection between the set of names and aliases for each group in each taxonomy. If a group in ATT&CK shares a name or alias with a group in Malpedia, we merge them by performing the union of their sets. After merging, we assign a unique name to each normalized group. We use the name in ATT&CK by default, and the name in Malpedia if the group is not in ATT&CK. The normalization process identified 145 groups common to both datasets, seven groups unique to ATT&CK, 655 groups found exclusively in Malpedia, and 807 groups in the union of both datasets. We perform a similar normalization for software. We first normalize all software names and aliases by removing common prefixes (e.g., trojan, win, apk, elf) and replacing special characters (e.g., _rat to rat). Then we merge software entries that share at least one normalized name. The normalization identifies 498 software that appear in both datasets, 2,869 only present in Malpedia, 296 only present in ATT&CK, and 3,663 in the union of both datasets.

Table 1 presents the overlap and union of report URLs, their FQDNs, and the SHA-256 of the downloaded reports. We find only 522 report URLs shared between ATT&CK and Malpedia, resulting in a low Jaccard Index of 3.2%, indicating minimal overlap in referenced sources. The overlap based on actual report content is even smaller, only 80 reports have identical SHA-256s, yielding a Jaccard Index of 0.5%. This discrepancy arises because downloading the same URL multiple times, particularly for HTML pages, can produce different files due to non-deterministic content, such as dynamic metadata or embedded advertisements. Overall, the overlap between the datasets is relatively low, with Malpedia providing a much broader view of the threat landscape. This disparity may be partly due to ATT&CK accepting contributions only from selected entities, which restricts the number of included threat reports. However, this selective approach contributes to under-reporting. To address this limitation, we build group profiles by combining group and software information from the union of both datasets. Note that technique information is only available from ATT&CK.

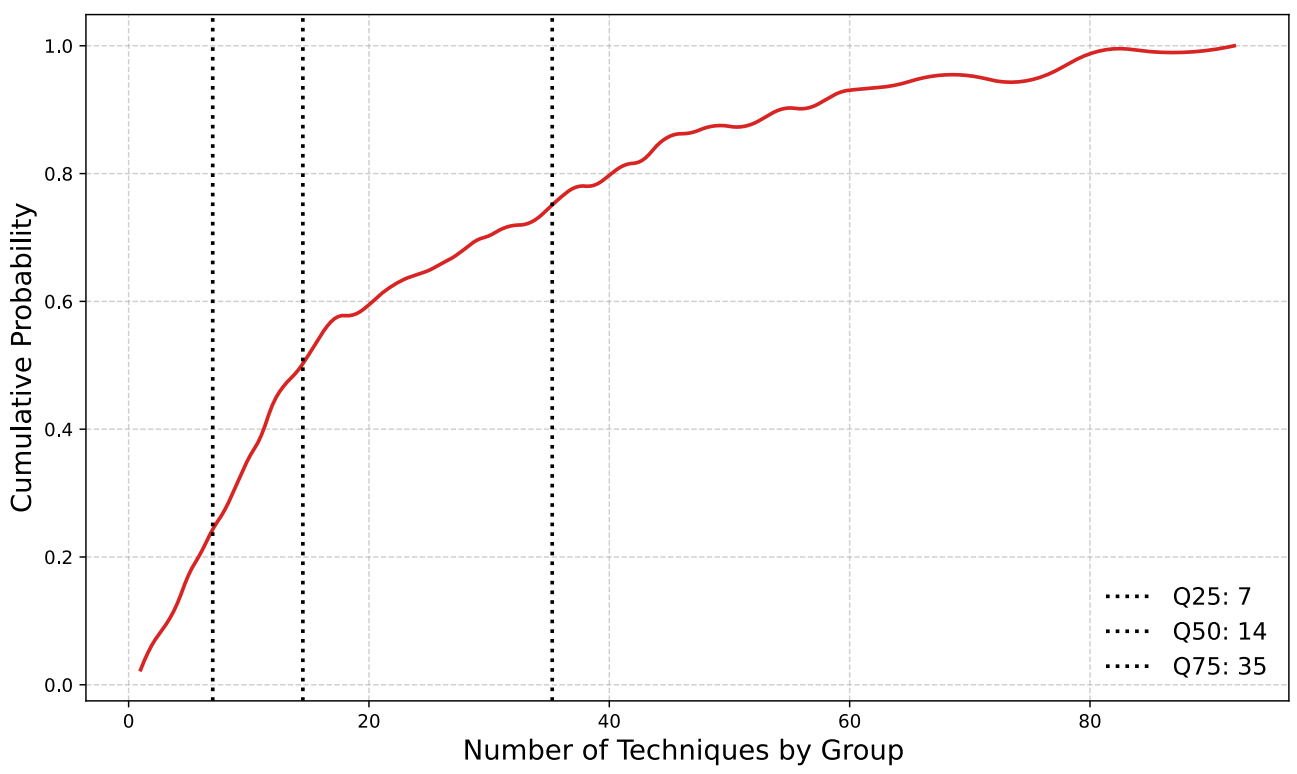

**Figure 1: CDF of the number of techniques used per group: 25% of groups use 7 or fewer techniques (Q25), 50% use 14 or fewer (Q50), and 75% use 35 or fewer (Q75).**

**Takeaway:** Malpedia provides a larger coverage of the threat landscape, including 5.2 times more groups and 4.2 times more software than ATT&CK. While not a strict superset of ATT&CK, Malpedia covers 95.4% of ATT&CK's groups and 62.7% of its software. Combining both datasets increases the overall coverage of the threat landscape.

# 3 Group Profiles in the Datasets

In this section, we first address RQ1 by quantifying the proportion of threat groups in ATT&CK and Malpedia that have group-specific behaviors. We then address RQ2 by evaluating whether combining the datasets improves the group profiles.

## 3.1 Technique Profiles

In ATT&CK, the association of techniques to groups is provided as three separate group spreadsheets, one per domain. We combine the three group spreadsheets to obtain the set of techniques associated with each group, which we term the group's *technique profile*.

We first measure the size of the technique profiles. Figure 1 shows the cumulative distribution function (CDF) of techniques per group. On average, each group uses 23.2 techniques. 38 (25%) have at most seven techniques, 76 (50%) have between 7 and 36, and 38 (25%) have more than 35 techniques. The Lazarus Group (G0032) has the highest number of techniques, with 92 techniques. Four groups have no associated techniques and therefore cannot be identified through their TTPs. We next examine whether the remaining 148 groups contain group-specific techniques.

To this end, we build a mapping from each technique and subtechnique to the threat groups that use them. Among the 839 techniques cataloged in the ATT&CK framework, 388 (46.3%) have not been associated with any group, 147 (17.5%) are linked to a single group, 287 (34.2%) are associated with 2–37 groups, and 17 (2.0%) are used by at least one quarter (38) of all groups. The fact that 388 (46.3%) of all techniques in ATT&CK are not associated with any group raises concerns about coverage, as these techniques were presumably added to the taxonomy based on observed adversary behavior, yet remain unlinked to any known group.

**Generic techniques.** We call techniques used by many groups generic, as their presence in a protected environment offers limited value in distinguishing specific adversaries. Table 2 lists the top 10 techniques by number of groups. The most common technique is *Malicious File* (T1204.002) used by 79 groups where adversaries rely on users opening a malicious file, followed by *Ingress Tool Transfer* (T1105, 76 groups) where adversaries transfer tools or files from an external system into a compromised environment, and *Spearphishing Attachment* (T1566.001, 72 groups) where emails with a malicious attachment are used as an initial compromise vector.

In addition, we analyze which pairs of techniques tend to occur together. For this we compute the co-occurrence rate

$$\frac{|G_A \cap G_B|}{max(|G_A|,|G_B|)}$$

where $G_A$ and $G_B$ are the sets of groups using techniques $A$ and $B$, respectively. We find five pairs with a co-occurrence rate of at least 0.75. The highest rates are 0.951 between *Malicious Link* (T1204.001) and *Spearphishing Link* (T1566.002), followed by 0.886 for *Malicious File* (T1204.002) and *Spearphishing Attachment* (T1566.001). The four techniques in these two pairs are generic, each used by at least one quarter of the groups, and are also semantically related, where a spearphishing link is a type of malicious link, and a spearphishing attachment is a malicious file that the user is lured to open.

**Technique profile similarity.** We compute the similarity of the technique profiles of each pair of groups using the Jaccard Index, after removing the top ten generic techniques in Table 2. The mean Jaccard Index is 0.04, the median is 0.02, and the maximum Jaccard Index is 0.50. Most pairs have low similarity scores, suggesting that their technique profile are unique. However, this may be due to a large space of techniques and limited coverage in ATT&CK.

Figure 2 shows a heatmap of the 12 groups with the highest Jaccard Index pairs (i.e., at least 0.33). These pairs are due to a small number of techniques in the profiles. For example, Scarlet Mimic (G0029) and Ferocious Kitten (G0137) each have two non-generic techniques, and they share one of them (*Masquerading: Right-to-Left Override*), resulting in a 0.50 Jaccard Index. Because there are few non-generic techniques, even a single shared technique gives a relatively high similarity score.

**Group-specific techniques.** We identify 147 (17.5%) group-specific techniques that are associated with a single group. Only 52 (34.2%) groups have group-specific techniques in ATT&CK. The mean number of group-specific techniques is 0.99. While most techniques are shared across groups, a few stand out for their distinctiveness, with the maximum number (16) of group-specific techniques observed in Windshift (G0112).

A key question is whether these group-specific techniques are due to limited coverage in ATT&CK, or if they truly represent capabilities developed or exclusively adopted by a single group. Table 3 provides examples of group-specific techniques. Some of these group-specific techniques appear indeed quite specific to their respective groups. For example, APT12 aka Numbered Panda (G0005) is the only group using *DNS Calculation* (T1568.003), where adversaries perform calculations on addresses returned in DNS results to determine which port and IP address to use for command and control. Conversely, some group-specific techniques may not be truly specific to their groups. For example, APT28 aka Fancy

**Table 2: Top 10 generic techniques, i.e., techniques used by the largest number of groups.**

| ID | Technique Name | Groups |
|---|---|---|
| T1204.002 | User Execution: Malicious File | 79 (9.8%) |
| T1105 | Ingress Tool Transfer | 76 (9.4%) |
| T1566.001 | Phishing: Spearphishing Attachment | 72 (8.9%) |
| T1059.001 | Command & Scripting: PowerShell | 69 (8.5%) |
| T1588.002 | Obtain Capabilities: Tool | 66 (8.2%) |
| T1059.003 | Command & Scripting: Win Cmd Shell | 60 (7.4%) |
| T1036.005 | Masquerading: Match Legitimate Name or Location | 50 (6.2%) |
| T1547.001 | Boot or Logon Autostart Execution: Registry/Startup Folder | 50 (6.2%) |
| T1071.001 | Application Layer Proocol: Web Protocols | 47 (5.8%) |
| T1082 | System Information Discovery | 46 (5.7%) |

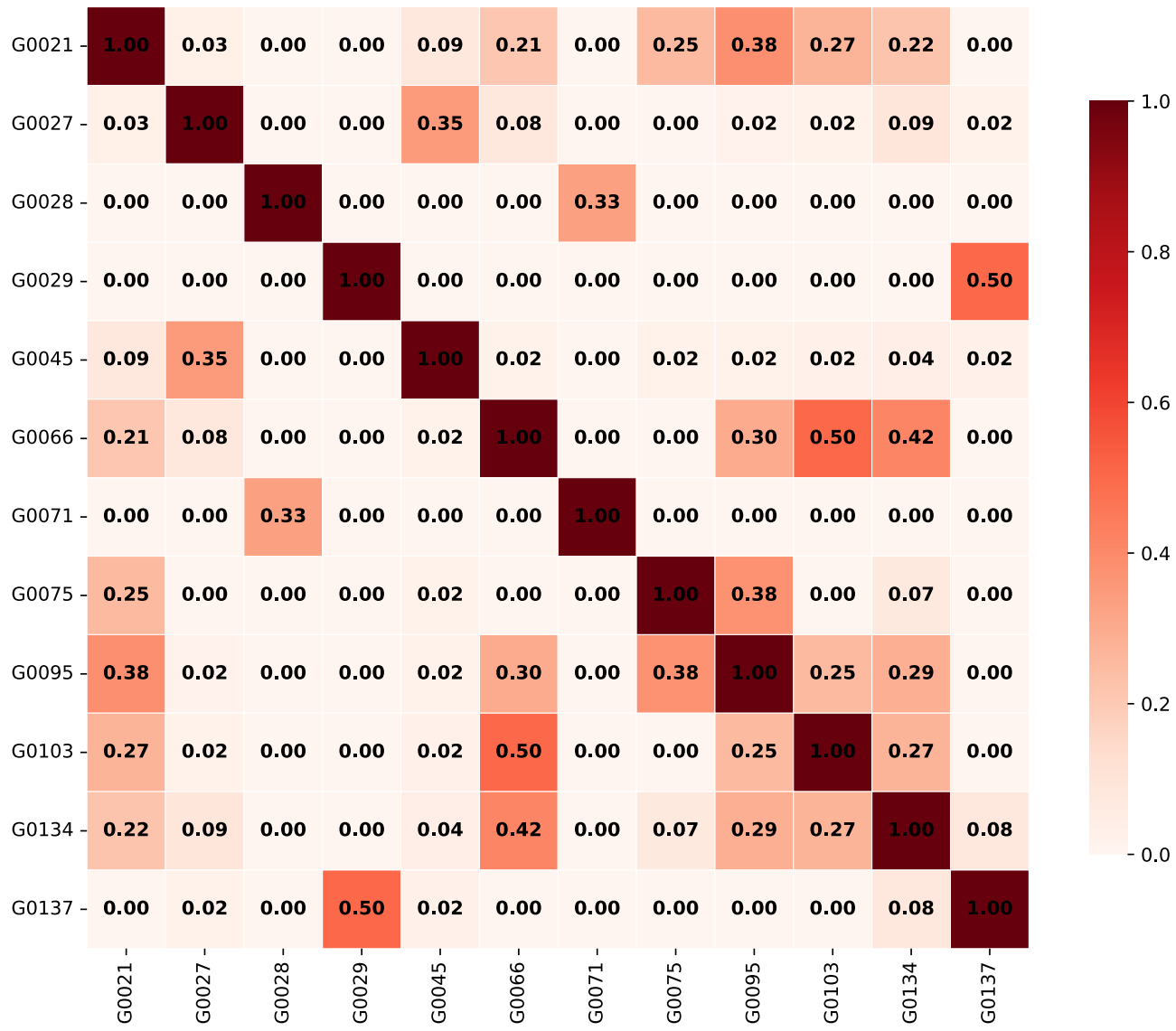

**Figure 2: Jaccard Index between the 12 most similar groups after removing the top ten generic techniques. The mean Jaccard Index across all groups is 0.04, the median is 0.02, and the maximum Jaccard Index is 0.50.**

Bear (G0007) is the only group associated with *Network Denial of Service* (T1498), a fairly common attack technique likely to be used by many groups, suggesting this uniqueness may reflect limited coverage rather than actual exclusivity.

**Beyond single group-specific techniques.** We further examine if there may be pairs or triplets of techniques that are specific to a group. We identify an additional 63 groups with group-specific technique pairs, whereas no additional groups have group-specific technique triplets. However, most pairs do not seem truly distinctive. For instance, APT-C-23 (G1028) is the only group with the pair *Match Legitimate Name or Location* (T1655.001) and *Phishing in Mobile* (T1660). However, this is likely due to most groups being associated with the more common *Phishing* (T1566) technique that is not specific to the mobile domain. In another case, Malteiro (G1026) exhibits two group-specific technique pairs: *Malicious File* (T1204.002) and *Financial Theft* (T1657), and *Security Software Discovery* (T1518.001) and *Financial Theft* (T1657). These largely reflect the group's financial motivation rather than a behavior specific to the group, i.e., other groups involved in financial theft may also exhibit those behavior pairs. Furthermore, using pairs of techniques to attribute a group requires both techniques to be observed in the compromised network, which makes evasion easier if adversaries vary just one technique. Overall, technique pairs appear less robust as signatures and would require human vetting to ensure they truly represent distinctive group behaviors. While manual validation may also be needed for single group-specific techniques, the need for vetting increases as the tuple size grows.

**Table 3: Examples of group-specific techniques, some groups have multiple group-specific techniques.**

| Group Name | Technique ID | Technique Name |
|---|---|---|
| APT12 | T1568.003 | DNS Calculation |
| APT28 | T1550.001<br>T1546.015<br>T1001.001<br>T1137.002<br>T1211<br>T1498 | Application Access Token<br>Component Object Model Hijacking<br>Junk Data<br>Office Test<br>Exploitation for Defense Evasion<br>Network Denial of Service |
| APT32 | T1552.002<br>T1564.004 | Credentials in Registry<br>NTFS File Attributes |
| APT37 | T1123 | Audio Capture |
| APT38 | T1562.003<br>T1565.003<br>T1565.001<br>T1565.002 | Impair Command History Logging<br>Runtime Data Manipulation<br>Stored Data Manipulation<br>Transmitted Data Manipulation |
| APT39 | T1546.010<br>T1059.010<br>T1056 | AppInit DLLs<br>AutoHotKey & AutoIT<br>Input Capture |
| APT41 | T1596.005 | Scan Databases |
| APT5 | T1554 | Compromise Host Software Binary |
| Axiom | T1563.002<br>T1001.002<br>T1553 | RDP Hijacking<br>Steganography<br>Subvert Trust Controls |
| Chimera | T1110.004<br>T1556.001 | Credential Stuffing<br>Domain Controller Authentication |
| Cobalt Group | T1218.008 | Odbcconf |
| DarkVishnya | T1200 | Hardware Additions |
| Darkhotel | T1497 | Virtualization/Sandbox Evasion |

**Unsupervised clustering.** We applied unsupervised clustering to identify ATT&CK groups using similar techniques. We use the Hierarchical Density-Based Clustering (HDBSCAN) [8] because it can find clusters of arbitrary shape and does not require prior knowledge of the number of clusters. It identified two clusters, one with 102 groups and another with three, marking the remaining groups as noise. The large cluster highlights how generic techniques used by multiple groups can make groups look similar.

**Takeaway:** Only 52 groups (34.2%) have group-specific techniques. However, other groups may still be distinguishable by group-specific technique combinations, as seen by the low mean Jaccard Index of 0.06. Under-reporting remains a concern, as only 53.7% of ATT&CK techniques are observed in group profiles, and some seemingly group-specific techniques may not be truly specific.

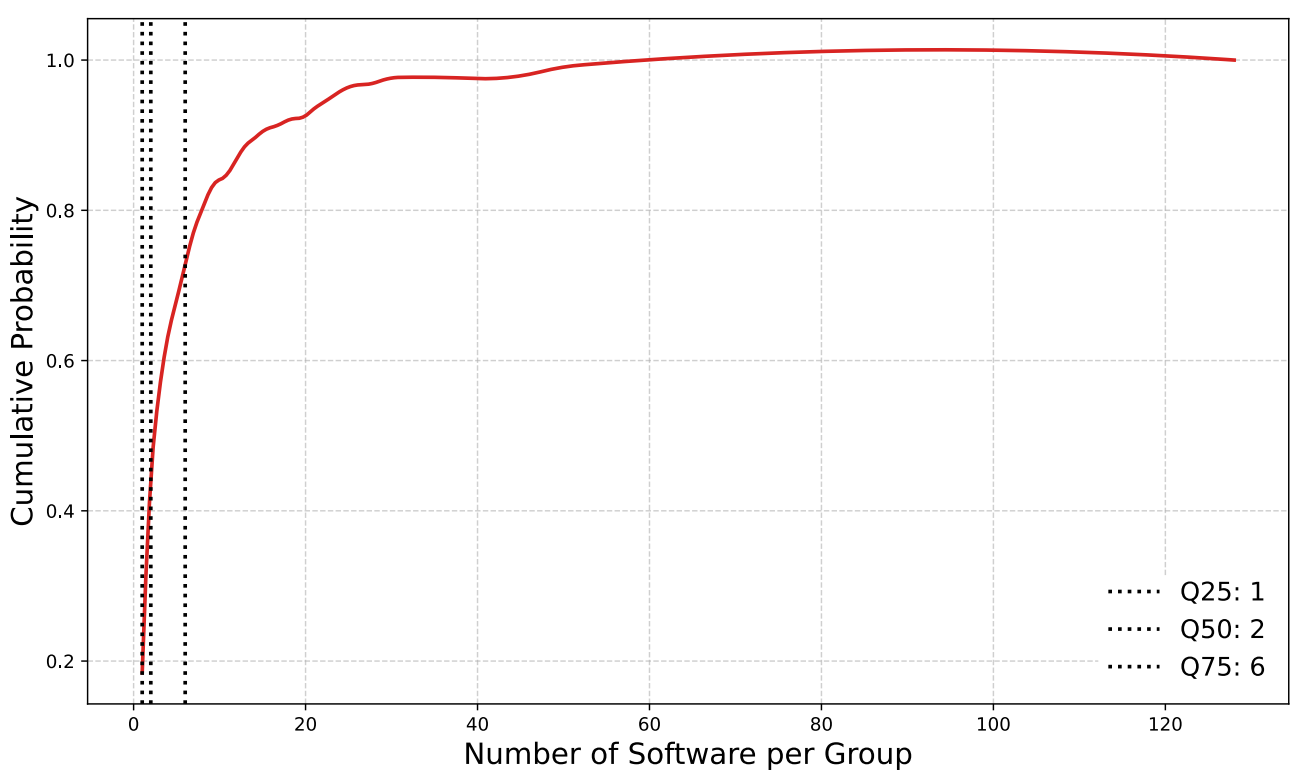

**Figure 3: CDF of the number of software per group. 25% of groups used one or fewer software (Q25). 50% of groups used 2 or fewer software (Q50). Most groups used a relatively small number of software, with 75% using 6 or fewer (Q75).**

## 3.2 Software Profiles

In this section, we explore whether the software used by each group identifies it. For each group, we build software profiles using the sets of normalized software names associated with the group in each dataset and from their union.

We first examine each dataset separately. Of the 794 software in ATT&CK, 509 (64.1%) are associated with at least one group. For Malpedia, the fraction is significantly smaller, where out of 3,367 software, only 963 (28.6%) are associated to at least one group. Software not associated to groups typically corresponds to non-APT malware. For example, the Conficker worm [31] and the Babuk ransomware [30] each appear in both ATT&CK and Malpedia and are not associated to groups in either dataset. The lower ratio of software associated to groups in Malpedia is likely due to Malpedia's larger coverage of non-APT malware.

The fraction of groups with a non-empty software profile is also larger in ATT&CK where 138 out of 152 (90.8%) groups have associated software compared to 220 (27.5%) out of 800 groups in Malpedia. However, Malpedia contains 16.9 times as many threat reports as ATT&CK, providing significantly more data for building software profiles. This difference arises because many lesser-known groups have few reports to support comprehensive profiling.

Next, we examine the unified software profiles. Out of the total 807 groups in both datasets, 264 (32.7%) groups have a non-empty software profile. Thus, over two-thirds of the groups cannot be identified by their associated software. Figure 3 shows the CDF of software per group for groups with at least one software. On average, each group uses 6.2 software, but most groups use a relatively small number of software, with 25% of the groups having only one associated software and 75% using six or fewer. The APT38 group (G0082) has the highest number of software samples, with 120.

**Generic software.** We call the software used by many groups generic, as their detection offers limited value in distinguishing specific adversaries. Table 4 lists the top 10 software by the number of groups where the software appears in the group's profile. All 10 software appear in ATT&CK, while only five are present in Malpedia. The ones missing from Malpedia include four OS tools (PsExec, Net, ipconfig, Ping) and the open-source Empire remote administration and post-exploitation framework [11]. Of these ten software entries, seven are classified as tools in ATT&CK and three as malware. Malpedia does not provide such classification. Among the three labeled as malware in ATT&CK, Cobalt Strike should arguably be categorized as a tool, as it is a commercial penetration-testing package [14], whereas PlugX and PoisonIvy are remote administration tools (RATs) commonly available on underground markets. In summary, generic software typically refers to tools and malware kits that are either commercially sold or widely accessible.

**Table 4: Top 10 generic software by number of groups using them, their type in ATT&CK (🔧=Tool or ⚙=Malware), whether they are in Malpedia (✓), and the number and percentage of groups using them.**

| ID | Name | ATT&CK Type | Malpedia | Groups |
|---|---|---|---|---|
| S0002 | Mimikatz | 🔧 | ✓ | 46 (5.7%) |
| S0029 | PsExec | 🔧 | ✗ | 31 (3.8%) |
| S0039 | Net | 🔧 | ✗ | 30 (3.7%) |
| S0154 | Cobalt Strike | ⚙ | ✓ | 26 (3.2%) |
| S0013 | PlugX | ⚙ | ✓ | 25 (3.1%) |
| S0363 | Empire | 🔧 | ✗ | 15 (1.8%) |
| S0012 | PoisonIvy | ⚙ | ✓ | 14 (1.7%) |
| S0100 | ipconfig | 🔧 | ✗ | 13 (1.6%) |
| S0097 | Ping | 🔧 | ✗ | 13 (1.6%) |
| S0349 | LaZagne | 🔧 | ✓ | 12 (1.5%) |

We identify 88 software tools that are not marked as such in ATT&CK but are used by multiple groups. These likely correspond to publicly available malware kits sold or shared on underground forums. In addition to Cobalt Strike, PlugX, and PoisonIvy (see Table 4), other commonly reused malware include gh0st RAT (used by 10 groups), China Chopper web shell (8 groups), 8.t dropper (8 groups), njRAT (7 groups), and ShadowPad (7 groups). Of these, gh0st is open-source [55], njRAT's source code was leaked [12], China Chopper is publicly available [9], and both 8.t and ShadowPad have been reported to be privately shared among Chinese threat groups [26, 52]. In some cases, the groups using the same software may be related. For instance, Bistromath is reported by Malpedia to be used by both Lazarus Group and Silent Chollima, the latter being a subsidiary of Lazarus [13].

**Group-specific software.** We term software as group-specific if it has only been associated with one group and is not classified as a tool in ATT&CK. We exclude tools, even when associated with a single group, because they can be readily adopted by others in the future, thereby providing weak attribution. Of the 3,663 software across both datasets, 952 (26.0%) are associated with a single group, making them group-specific software. The detection of group-specific software may allow attributing the group behind an attack. Of the 807 total groups, 213 (26.3%) have at least one group-specific software associated with them. Among these 213 groups, the mean and median number of group-specific software are 4.5 and 1.0, respectively, and 130 groups (16.1%) have only group-specific software. Some threat groups develop many custom tools for their attacks, for example, APT38 (G0082) has the highest number, with 99 group-specific software.

**Beyond single group-specific software.** We also analyze group-specific software pairs. Beyond the 111 groups with group-specific software in ATT&CK, we identify 11 additional groups with no group-specific software but with group-specific software pairs. However, similar to what we observe with group-specific technique pairs, group-specific software pairs do not look truly specific to the groups. For example, DarkVishnya (G0105) is the only group using both PsExec (S0029) and Winexe (S0191). However, both PsExec and Winexe are common administrative tools used by multiple groups, so the combination of the two tools does not appear to be a robust signature for that group. Overall, all 11 groups with group-specific software pairs include a benign tool in each pair, indicating that the software pairs are not robust as behavioral signatures.

**Takeaway:** In ATT&CK, 111 groups (73.0%) have group-specific software, whereas in Malpedia, only 192 groups (24.0%) have that. Combining both datasets increases the number of groups to 213 (26.3%), which still remains relatively low. Among the 264 groups with non-empty software profiles, the median number of software is 2, indicating that most threat groups operate with limited toolsets. However, some groups maintain extensive custom toolsets, for example, APT38 has 99 group-specific software.

## 3.3 Combining Group Profiles

Table 5 summarizes the number of groups with non-empty profiles and those with at least one group-specific entry across all different profile-building methods we examined. The first three rows correspond to the profiles we discussed in this section, based on techniques only, software only, and the combination of both.

To answer RQ1, our results show that identifying groups using techniques is challenging, as only 52 (34.2%) groups in ATT&CK have group-specific techniques. Identification is easier using software, with 111 (73.0%) of ATT&CK groups having group-specific software. However, this percentage is significantly lower in Malpedia, where only 192 (24.0%) of groups have group-specific software. The lower ratio in Malpedia likely reflects its focus on a subset of high-profile groups, while many smaller groups have too few reports to build robust profiles. Joint profiles combining both techniques and software improve identification in ATT&CK by 12.6 percentage points, increasing the number of groups with group-specific behaviors from 111 (73.0%) to 124 (81.6%).

To answer RQ2, the joint profiles built combining data from both datasets identify 236 (29.2%) groups with group-specific behaviors, compared to 124 groups using only ATT&CK and 192 using only Malpedia. Nonetheless, even when combining techniques and software, over 70% of groups do not have any group-specific behavior.

# 4 Extending the Group Profiles

So far, the group profiles we examined have included techniques and software from both ATT&CK and Malpedia. In this section, we address RQ3, i.e., whether we can extend the group profiles with additional data extracted from the downloaded threat reports. In Section 4.1, we build a vulnerability profile for each group with the vulnerabilities that the threat reports refer to as being used by the group in its attacks. Then, in Section 4.2, we discuss how we extend the technique profiles with additional technique identifiers extracted from the threat reports. This allows incorporating techniques mentioned from the threat reports in Malpedia and to examine how complete the technique extraction was in ATT&CK.

**Table 5: Summary of group profiles across different data sources and combinations. We report the number of groups with non-empty profiles and the subset with at least one group-specific behavior. The top section presents results for profiles built using only techniques, only software, and a combination of both, as analyzed in Section 3. The middle section shows results from profiles enriched with extracted CVE identifiers and additional techniques from downloaded threat reports, discussed in Section 4. The bottom row shows the most comprehensive profiles, combining all available behavioral indicators.**

| Data Source(s) for Group Profiles | ATT&CK | | Malpedia | | ATT&CK ∪ Malpedia | |
|---|---|---|---|---|---|---|
| | Non-Empty | w/Group-Specific | Non-Empty | w/Group-Specific | Non-Empty | w/Group-Specific |
| *Techniques* | 148 (97.4%) | 52 (34.2%) | - | - | 148 (18.3%) | 52 (6.4%) |
| *Software* | 138 (90.8%) | 111 (73.0%) | 220 (27.5%) | 192 (24.0%) | 264 (32.7%) | 213 (26.3%) |
| *Techniques* ∪ *Software* | 151 (99.3%) | 124 (81.6%) | 220 (27.5%) | 192 (24.0%) | 331 (41.0%) | 236 (29.2%) |
| *Vulnerabilities* | 86 (56.6%) | 48 (31.6%) | 261 (32.6%) | 112 (14.0%) | 277 (34.3%) | 119 (14.7%) |
| *Techniques++* | 149 (98.0%) | 60 (39.5%) | 204 (25.5%) | 69 (8.6%) | 242 (30.0%) | 68 (8.4%) |
| *Techniques++* ∪ *Software* ∪ *Vulnerabilities* | 152 (100%) | 128 (84.2%) | 391 (48.9%) | 265 (33.1%) | 418 (51.7%) | 291 (36.0%) |

## 4.1 Vulnerability Profiles

The selection of which vulnerabilities to exploit is largely group-specific since it depends on the expected software used by the targets and the exploits the group has access to. The observation of specific vulnerabilities being exploited in a monitored system could potentially be used to attribute the group behind an attack. To build the vulnerability profiles, we first extract CVE identifiers from the downloaded threat reports using the `iocsearcher` open-source tool [7]. We then assign CVEs to groups. ATT&CK reports are associated with a single threat group, whereas Malpedia reports may reference multiple groups. When a report lists multiple groups, it is unclear to which group the CVEs in the report should be assigned. Therefore, we extract CVEs only from the 4,414 (29.4%) Malpedia reports that reference a single group, along with all 920 ATT&CK reports.

The left part of Table 7 summarizes the extraction of CVE identifiers from the downloaded threat reports. Among the 5,827 reports that we analyzed, 1,186 (20.3%) contain at least one CVE identifier for a total of 906 unique CVEs associated to 277 groups. Of the 807 groups, 277 (34.3%) have a non-empty vulnerability profile. The other 530 (65.7%) groups have no vulnerabilities that can be used to identify them. The mean CVEs per group is 8.9, and the maximum is 176 CVEs reported for APT28 aka Fancy Bear (G0007).

**Generic vulnerabilities.** Overall, there are 368 vulnerabilities used by at least two groups, 114 used by more than five groups, and 28 used by more than 10 groups. Table 6 lists the top 10 CVEs by the number of groups using them. These generic vulnerabilities target popular software, with Microsoft Office being the most frequently targeted one, with four vulnerabilities. Nine of the 10 vulnerabilities have publicly available proof-of-concept (PoC) exploits, either in the Exploit Database [36] or on GitHub. We did not find any PoC for CVE-2022-38028, which was a zero-day on the Windows Print Spooler used by Russian threat groups [16].

**Group-specific vulnerabilities.** Of the 906 CVEs identified in the reports, 538 (59.4%) are associated to a single group. We call these group-specific vulnerabilities. The *Vulnerabilities* row in Table 5 summarizes the generated profiles. Of the 807 groups, 119 (14.7%) have at least one group-specific CVE. Across these 119 groups, the mean and median vulnerabilities per group are 4.5 and 2.0, respectively. The maximum is for Gorgon (G0078), which uses 78 CVEs. Table 8 shows examples of group-specific CVEs.

**Table 6: Top 10 generic CVEs by number of groups using them, the software they target, and whether a proof-of-concept (PoC) exploit is publicly available (✓).**

| Vulnerability | Affected Software | PoC | Groups |
|---|---|---|---|
| CVE-2017-11882 | Microsoft Office | ✓ | 46 (5.7%) |
| CVE-2012-0158 | Microsoft Office | ✓ | 41 (5.1%) |
| CVE-2017-0199 | Microsoft Office | ✓ | 34 (4.2%) |
| CVE-2021-44228 | Apache Log4j | ✓ | 28 (3.5%) |
| CVE-2022-30190 | Microsoft Windows (MSDT) | ✓ | 25 (3.1%) |
| CVE-2022-26134 | Atlassian Confluence | ✓ | 21 (2.6%) |
| CVE-2018-0802 | Microsoft Office | ✓ | 20 (2.5%) |
| CVE-2022-38028 | Windows Print Spooler | ✗ | 17 (2.1%) |
| CVE-2023-38831 | RARLAB WinRAR | ✓ | 17 (2.1%) |
| CVE-2024-37085 | VMware ESXi | ✓ | 16 (2.0%) |

**Takeaway:** Only 119 (14.7%) groups have a group-specific vulnerability. The set of vulnerabilities exploited by a group is less specific than the set of software (27.8% of groups have group-specific software), likely because many groups focus on the same generic vulnerabilities affecting popular software, often with publicly available exploits. However, vulnerabilities tend to be more group-specific than techniques, as only 6.4% of groups have group-specific techniques.

## 4.2 Technique Identifiers in Reports

To extend the technique profiles, we look for *explicit mentions* of technique identifiers in the downloaded reports. It is important to note that threat reports may also include *implicit references* to techniques, such as stating that a rootkit was used without explicitly providing a technique identifier. We discuss the extraction of implicit references in Appendix A.

**Table 7: Summary of CVEs and technique identifiers extracted from the downloaded threat reports. For each dataset, we show the total number of reports analyzed, the number of reports containing at least one CVE, the number of unique CVEs extracted, and the number of threat groups associated with those CVEs. We provide analogous statistics for the technique identifiers.**

| Dataset | # of Reports | Reports w/CVE(s) | CVEs | Groups | Reports w/Technique(s) | Techniques | Groups |
|---|---|---|---|---|---|---|---|
| *ATT&CK* | 920 | 266 (29.0%) | 325 | 86 | 122 (13.2%) | 470 | 63 |
| *Malpedia* | 4,414 | 943 (21.3%) | 853 | 261 | 541 (12.2%) | 626 | 204 |
| *All* | 5,827 | 1,186 (20.3%) | 906 | 277 | 650 (11.1%) | 658 | 211 |

**Table 8: Examples of group-specific vulnerabilities and the software they target.**

| Group Name | Vulnerability | Affected Software |
|---|---|---|
| APT37 | CVE-2015-3636<br>CVE-2016-0147 | Linux Kernel<br>MSXML |
| Akira | CVE-2019-6693<br>CVE-2023-29336<br>CVE-2023-35078 | Fortinet FortiOS<br>Microsoft Windows<br>Ivanti Endpoint Manager |
| Kimsuky | CVE-2012-4873<br>CVE-2018-14745<br>CVE-2018-2628 | GNU Board<br>Samsung Galaxy<br>Oracle WebLogic Server |
| Gorgon | CVE-2015-7036<br>CVE-2019-8457<br>CVE-2019-8598 | Apple iOS<br>SQLite3<br>iOS, macOS |
| Sidewinder | CVE-2018-4876<br>CVE-2018-7445<br>CVE-2019-2215 | Adobe Experience Manager<br>MicroTik RouterOS<br>Google Android |
| Scattered Spider | CVE-2015-2291<br>CVE-2021-35464<br>CVE-2022-0001 | Ethernet driver on Windows<br>ForgeRock Acess Management<br>Intel Processors |
| Carbanak | CVE-2013-2463<br>CVE-2015-2426<br>CVE-2016-1010 | Oracle JRE<br>Microsoft Windows<br>Adobe Flash Player |

To identify technique identifiers, we use a regular expression provided by `iocsearcher` [7]. The technique identifiers, if present, are typically provided in a table at the end of the threat report, although they may also appear throughout the text.

The right part of Table 7 summarizes the extraction of technique identifiers from the downloaded threat reports. From the 4,414 Malpedia reports uniquely assigned to one group, we find 626 unique technique identifiers associated to 204 groups appearing in 541 (12.2%) reports. Of these 626 techniques, 248 are not associated with any groups in ATT&CK, i.e., are mentioned only in the Malpedia reports. This shows that focusing on a small set of reports leads to underreporting and that technique profiles extracted from ATT&CK are likely to miss techniques used by a group.

We further apply `iocsearcher` to the 920 reports downloaded from ATT&CK reference URLs and identify 470 unique technique identifiers from 122 (13.2%) reports. These 122 reports are associated with 63 groups. ATT&CK has 451 techniques associated with 152 groups, while we find a larger technique set (470) mentioned for 63 groups in 13.2% of the same reports. This indicates that ATT&CK contributors may not be systematic in extracting all references of techniques in the reports. Furthermore, we are only accounting for explicit references through technique identifiers. Threat reports may also include implicit references. However, extracting implicit references to techniques would likely reinforce the under-reporting trends we already observe. To recover the ATT&CK technique identifiers implicitly mentioned in threat reports, previous work has proposed using natural language processing (NLP) [1, 17, 43]. An alternative approach would be to use large language models (LLMs), which have demonstrated their flexibility in a number of security-related tasks involving natural language texts [10, 53]. Appendix A discusses our initial work to automate the extraction of implicit references using LLMs.

*Techniques++* in Table 5 captures the techniques profiles extended with the technique identifiers extracted from the threat reports. It shows that we can identify 69 groups using group-specific techniques from Malpedia threat reports, compensating for Malpedia's lack of techniques. Notably, when we extract technique identifiers directly from the ATT&CK reports and combine them with the existing ATT&CK technique profiles, the number of groups with group-specific techniques increases from 52 (34.2%) to 60 (39.5%), even though we analyze the same set of threat reports. This highlights that the manual extraction of techniques by ATT&CK contributors is not always optimal.

**Takeaway:** The extraction of technique identifiers from threat reports increases the groups with group-specific techniques from 52 (6.4%) to 119 (14.7%), mostly due to additional techniques in the Malpedia reports. Comparing the technique identifiers extracted from reports with those indexed in ATT&CK shows that ATT&CK contributors may miss techniques, indicating a need for automation.

## 5 Discussion

In this section, we discuss the limitations of our approach, mainly due to partial coverage, the implications of our results, and potential improvements to the analyzed datasets.

### 5.1 Limitations due to Partial Coverage

A key observation of our work is that determining whether a behavior is truly specific to a threat group requires good visibility on other threat groups. Otherwise, group-specific behaviors may not be truly group-specific, but rather appear to be so due to gaps in the datasets. For example, APT28 was the only group in ATT&CK associated with *Network Denial of Service* (T1498), a fairly common attack technique used by many other groups. This likely happens due to under-reporting. As more threat reports are analyzed, behaviors previously considered specific to a threat group may be observed in other threat groups, making them no longer group-specific. Thus, comprehensive coverage of the threat ecosystem is essential for building accurate threat group profiles.

**Table 9: Entries published in the blogs of 11 cybersecurity vendors between the specified dates, the subset of those corresponding to threat reports analyzing a malware family or APT, and the fraction of those that appear in Malpedia.**

| Cybersecurity Vendor | Days (First → Last) | First Entry | Last Entry | Entries | Reports | Included in Malpedia |
|---|---|---|---|---|---|---|
| Avast | 351 | 2024-02-28 | 2025-02-13 | 10 | 4 | 1 (25.0%) |
| BitDefender | 320 | 2024-09-12 | 2025-07-29 | 22 | 9 | 2 (22.2%) |
| Cisco-Talos | 34 | 2025-06-26 | 2025-07-30 | 17 | 2 | 0 ( 0.0%) |
| Eset-WeLiveSecurity | 220 | 2024-12-16 | 2025-07-24 | 19 | 13 | 8 (61.5%) |
| Fortinet | 32 | 2025-06-23 | 2025-07-25 | 12 | 7 | 0 ( 0.0%) |
| GenDigital | 46 | 2025-06-13 | 2025-07-29 | 11 | 3 | 0 ( 0.0%) |
| Kaspersky-SecureList | 51 | 2025-06-09 | 2025-07-30 | 12 | 8 | 2 (25.0%) |
| PaloAlto-Unit42 | 50 | 2025-06-10 | 2025-07-30 | 21 | 8 | 2 (25.0%) |
| TrendMicro | 130 | 2025-03-21 | 2025-07-29 | 54 | 18 | 8 (44.4%) |
| ZScaler | 425 | 2024-05-30 | 2025-07-29 | 51 | 31 | 18 (58.1%) |
| Zimperium | 62 | 2025-05-29 | 2025-07-30 | 12 | 4 | 0 ( 0.0%) |
| **All** | 518 | 2024-02-28 | 2025-07-30 | 241 | 107 | 41 (38.3%) |

We have identified three main factors affecting coverage that may limit the generalizability of our results: (1) the type of datasets used, (2) the dataset coverage, i.e., how completely they capture the entire threat ecosystem, and (3) the approach used by the datasets to extract behaviors from the threat reports. We discuss each of these factors below:

**Dataset selection.** Our work analyzes the feasibility and challenges of building threat group behavioral profiles from two public datasets (ATT&CK and Malpedia), rather than the general viability of behavioral profiling of threat groups. Private CTI sources (e.g., commercial feeds, private threat exchanges) may cover additional threat groups, offer more detailed threat group analysis, and include behaviors not present in public reports. There are two main reasons we focus on public CTI. First, the subscription for a leading commercial CTI feed usually ranges in the hundreds of thousands of dollars per year [5], making access to even a single commercial feed prohibitive for most research groups. Second, previous work has analyzed CTI feeds providing indicators of compromise (IOCs) such as IP addresses, domain names, and file hashes [5, 62]. Their results show that both public [62] and private [5] IOC feeds are affected by similar low coverage limitations, i.e., any two feeds (regardless of whether public or private) have little IOC overlap. Their findings hint that each vendor may only focus on a subset of threats. Similarly, it is possible that each vendor focuses on specific threat groups, with little overlap across vendors except for the most prominent groups, Thus, no single vendor likely provides sufficient coverage to build accurate group profiles. For this reason, we use public datasets like ATT&CK and Malpedia, which aggregate threat reports from multiple vendors.

A limitation introduced by our focus on public CTI data is that cybersecurity vendors may omit relevant threat group behavioral markers from their public reports. If the removal aims to keep such data commercially relevant, the removed behavioral markers may still appear in the vendor's commercial feeds. On the other hand, if they are removed to protect the privacy and security posture of their clients, they will likely be removed not only from public reports, but also from other sources (e.g., commercial feeds, private threat exchanges) for the same reason.

**Dataset coverage.** Even large aggregators like Malpedia may only cover a subset of all publicly available threat reports. To estimate Malpedia's coverage, we collected daily the entries in the RSS feeds of 11 cybersecurity vendors between July 24 and July 30, 2025. Each RSS entry corresponds to a post in the vendor's blog with a URL, a title, an author, and a publication date. While our collection lasted one week, each feed provides a configurable number of the last published entries. Thus, the 241 blog entries collected were published over time periods ranging from 32 days for Fortinet up to 425 days for ZScaler, as summarized in Table 9.

The vendors publish diverse posts in their blogs, including advertisements of their products, periodic summaries of the threat landscape, security recommendations, and technical analyses of threats. We manually examined the 241 entries, identifying 107 (44.4%) that we believe fall within Malpedia's scope because they capture the analysis of malware or APTs. Of those 107 threat reports, 41 (38.3%) were included in the Malpedia BibTeX file as of August 6, 2025. We also checked the latest Malpedia BibTeX file from November 30, 2025, but the coverage remained the same.

This experiment shows that aggregators like Malpedia may cover only a fraction of the publicly available CTI data, which limits the completeness of the group profiles generated from such data. Automating the collection of public CTI from a diverse set of sources [7] could help improve the coverage and representativeness of public threat intelligence knowledge bases.

**Behavior extraction.** Even when a threat report is part of a knowledge base, not all the behaviors it describes may be included in the knowledge base. There are two main reasons for this. First, some behaviors may not be extracted. For example, Malpedia does not include techniques in its group profiles, and neither Malpedia nor ATT&CK include exploited vulnerabilities. To address this issue, we extended the group profiles with techniques and vulnerabilities directly extracted from the threat reports (see Section 4).

Second, the methodology used to extract behaviors from threat reports also has an impact. Currently, such extraction is mainly manual, causing some behaviors mentioned in the reports to be missed. While ATT&CK had 451 techniques associated with 152 groups, we could extract a larger technique set (470) from explicit technique references in a subset of 13.2% of the ATT&CK reports. This suggests that the manual extraction process used by ATT&CK contributors is not exhaustive, motivating the need for automated approaches, for example using NLP techniques [1, 17, 43], or LLMs. We provide our initial results on using LLMs to extract implicit technique references from threat reports in Appendix A.

## 5.2 Other Limitations

**Temporal evolution.** As the features and tactics [22, 24] of a group evolve, and as new analyses of the group's activities are published, the profile of a group changes over time. Unfortunately, most groups have too few reports, which limits our ability to analyze the evolution of their profiles. Such analysis is only possible for the few groups with multiple reports over several years. For the interested reader, Appendix B illustrates the evolution of the profile for the Lazarus Group (G0032) that has the most reports.

**Beyond group-specific behaviors.** We have explored group-specific single behaviors, as well as group-specific behavioral pairs and triplets. However, our approach may overlook groups characterized by combinations of non-exclusive behaviors. An alternative would be using supervised machine learning (ML) classifiers, which we did not evaluate due to the lack of reliable ground-truth labels required for training and evaluating them.

**Additional behaviors.** Group profiles could be further extended with other behaviors in threat reports, such as payment services used to receive victim payments, communication channels through which victims and attackers interact, and the textual content shown to victims (e.g., emails, ransom notes).

**Naming.** Vendors often use aliases to refer to the same malware family [25, 33, 51] and threat group [46, 57]. To address this issue, we have applied a normalization strategy to align group and software names (see Section 2.2). However, some of our mappings could be incorrect. We publicly release our mappings of group and software names as part of our artifact for future reference.

## 5.3 Implications of Results

Group-specific behaviors can be used as behavioral signatures to identify the threat group responsible for an attack, if they are distinct enough [39]. Out of 807 threat groups analyzed, 64% of threat groups do not have group-specific behaviors, or we failed to identify them due to limited coverage. However, a non-trivial 36% threat groups had group-specific techniques, software, or vulnerabilities.

**Building more complete profiles.** Increasing coverage is fundamental for building accurate threat group profiles that enable identifying truly group-specific behaviors, which is critical given the implications of threat group attribution. For example, misattributing an attack on country A as having been launched by country B, when A and B were at war recently, could start a new war. Coverage can increase by adding new sources of threat reports and by using automated extraction approaches [1, 17, 43]. In addition, our work hints at the need to increase the diversity of threat groups analyzed. Notably, going from 152 groups in ATT&CK to 800 groups in Malpedia reduces the ratio of groups with distinct behaviors from 84.2% to 33.1% because ATT&CK focuses on the most popular threat groups, which have received significant attention. Instead, Malpedia contains a long tail of threat groups that have received much less attention, with only 51.7% of the Malpedia groups having a non-empty behavioral profile. Thus, it is fundamental that attention is not only focused on a few highly impactful threat groups, but a larger number of groups receive enough attention.

**Most-specific behaviors.** The software used by a group is the most distinctive behavior in both ATT&CK (73.0% of groups have group-specific software) and Malpedia (24.0%), while vulnerabilities are least distinctive in ATT&CK (31.6%) and techniques in Malpedia (8.6%). Removing generic behaviors is important to build accurate behavioral signatures. Identifying generic software is the most straightforward, as it corresponds to publicly available software such as tools, commercial software, and malware kits sold underground. Generic vulnerabilities tend to have public exploits. Identifying generic techniques is the hardest as it requires ranking techniques by the number of groups that have used them, which increases over time.

**Legal use of behavior-based attribution.** A key challenge with behavior-based attribution is that just because a behavior was used only by one threat group in the past, there is no guarantee the group is still using that behavior, and that other groups have not adopted the behavior. Behavioral profiles provide circumstantial evidence with significant risk of misattribution, especially since different groups may exhibit similar behaviors (e.g., using publicly available software or common techniques). For these reasons, behavioral profiles may face an uphill battle in court, being susceptible to challenges from opposing experts that dispute the profile generation methodology or the conclusions drawn from them.

Furthermore, threat groups are pseudo-anonymous. To initiate legal proceedings, the individuals or legal entities behind a threat group must be identified. On a positive note, behavioral profiles may provide enough plausible evidence for a judge to authorize further evidence collection.

## 5.4 Dataset Improvements

Our analysis reveals the following potential dataset improvements:

**Software categories.** The classification of software into tools and malware in ATT&CK is useful to filter generic software, but it is missing in Malpedia. Furthermore, it is not clear where malicious kits should be placed, possibly indicating the need for a third category. For tools, it may be possible to leverage existing efforts like Common Platform Enumerations (CPE) [35] to avoid yet another taxonomy. We also observe some likely misclassified software, e.g., Cobalt Strike is arguably a tool rather than malware.

**ATT&CK domains.** The split of techniques into domains used by ATT&CK seems quite arbitrary as techniques may apply to different domains, albeit with different implementations. Indeed, there are instances of equally named techniques in different domains such as *Rootkit* (T1014) in the Enterprise domain and *Rootkit* (T0851) in the ICS domain. While the latter includes in the description references to firmware rootkits and Stuxnet, having two equally named techniques is confusing and likely unnecessary. The Enterprise domain also includes *Pre-OS Boot: System Firmware* (T1542.001) for capturing adversaries modifying system firmware to persist on systems, which seems to describe a firmware rootkit. The split into domains also complicates usage, as three technique taxonomies, one per domain, need to be considered. For example, given an observation of a rootkit on a device, different security vendors may assign it any combination of the above three techniques. This makes it tempting to focus on the Enterprise domain, comprising 76% of all techniques and sub-techniques. A more unified taxonomy, with domains encoded as an attribute, could streamline analysis.

**Missing reports.** Knowledge bases identify threat reports by their URLs, which may become unavailable over time, e.g., if the vendor goes bankrupt or is acquired by another vendor. In our data collection, roughly 5% of report URLs pointed to inactive resources. To avoid losing reports, knowledge bases should send the report URLs to archives such as the Wayback Machine [18] or APTnotes [3].

## 6 Related Work

Our research relates to the following prior CTI research:

**Knowledge bases.** Previous work has presented the design of the two knowledge bases we use [40, 58]. Other work has analyzed the usage of ATT&CK by systematically reviewing literature on its applications [19, 47, 48]. Frequently, research uses the knowledge bases simply as a source of threat reports from where IOCs can be extracted [7, 21]. Our work differs in that it measures the utility of ATT&CK and Malpedia for the specific case of adversary profiling.

**Application-oriented studies.** Several studies have examined the use of ATT&CK across different cybersecurity contexts. Oosthoek et al. [37] employed ATT&CK to map sandbox evasion techniques across 951 Windows malware families, offering insight into both commonly used and increasingly adopted techniques in recent years. Virkud et al. [63] evaluate the ATT&CK framework in commercial endpoint detection products and assess its effectiveness as a security evaluation metric. They find that while these products typically cover between 48%–55% of ATT&CK techniques, much of this coverage consists of low-risk or less impactful rules. Their findings suggest that although ATT&CK is increasingly used to assess threat readiness, reported coverage frequently fails to reflect actual detection capabilities in real-world scenarios. In another line of work, Rahman et al. [42] investigate challenges in implementing security controls (e.g., strong password policies) against ATT&CK techniques. In simultaneous and independent work, Horst et al. [61] examine the role of low-level IOCs (e.g., domains) and high-level IOCs (e.g., TTPs) in ransomware attribution. They use a mixed-methods approach, combining interviews with 15 ransomware attribution experts and the analysis of 27 incident reports from two sources. They show that experts rely more frequently on low-level IOCs for attribution than on high-level IOCs, which they regard as too generic. Our results align with theirs in raising concerns about the use of behavioral traits for attribution. But our approaches are quite different: Horst et al. examine 16 ransomware groups, while we examine 807 threat groups covering different types of adversaries (e.g., APTs). In addition, we do not conduct interviews but instead analyze more than 15,000 threat reports from two popular knowledge bases. Most importantly, we measure for the first time the fraction of threat groups with group-specific behaviors.

**Automated CTI extraction.** Husari et al. [17] made early efforts to automate the extraction of TTPs from threat intelligence reports, using a context-aware, rule-based approach to identify and extract threat actions from both structured and unstructured CTI sources. The extracted TTPs are standardized using the STIX [29] format, with the tool achieving over 82% precision and recall on a proprietary dataset. Extending this work, Alam et al. [1] employed machine learning for automated extraction of attack patterns and IOCs. Their framework further mapped the extracted behaviors to the standardized ATT&CK framework and organized them in a knowledge graph to facilitate predictive analysis. Complementing these extraction-focused efforts, Rahman et al. [41] analyzed 667 CTI reports from the ATT&CK framework to study the prevalence and co-occurrence of TTPs used in APT campaigns, providing insights into adversary patterns. Our work builds upon these approaches by combining threat intelligence data from both the ATT&CK framework and Malpedia. We examine techniques and vulnerability usage across adversary groups, offering insights into building more comprehensive threat group profiles.

## 7 Conclusion

Our work measured the distinctiveness of the threat group behavioral profiles using public knowledge bases, investigated how these profiles can be complemented with additional vulnerability and technique data, and, quite crucially, quantified to what extent the produced profiles support reliable threat group attribution. Out of 807 threat groups that we analyzed, 64% threat groups do not have group-specific behaviors, or we failed to identify them due to limited coverage. However, for 36% of the threat groups, we indeed identified group-specific techniques, software, or vulnerabilities, which could potentially be used as behavioral signatures to identify their attacks. Our work identified limited coverage in public CTI knowledge bases as a key challenge in building accurate behavioral profiles that can identify true group-specific behaviors, avoiding false positives. Consequently, improving coverage emerges as a clear avenue for the CTI community to build distinctive behavioral profiles and enable enhanced profile-based attack attribution.

## Acknowledgments

This work has been partially supported by MICIU/AEI/10.13039/501100011033 through grants PID2022-142290OB-I00 (ESPADA), CEX2024-001471, and RED2024-154240-T (EMACS). These projects are co-funded by the European Union EIE and NextGeneration EU/PRTR Funds. This work has also been supported by the Vienna Science and Technology Fund (WWTF) and the City of Vienna under the projects IoTIO [Grant ID: 10.47379/ICT19056] and BREADS [Grant ID: 10.47379/VRG23011], as well as SBA Research (SBA-K1 NGC), a COMET Center within the COMET – Competence Centers for Excellent Technologies Programme and funded by BMIMI, BMWET, and the federal state of Vienna. The COMET Programme is managed by FFG.

## References

[1] Md Tanvirul Alam, Dipkamal Bhusal, Youngja Park, and Nidhi Rastogi. 2023. Looking Beyond IoCs: Automatically Extracting Attack Patterns from External CTI. In *Proceedings of the 26th International Symposium on Research in Attacks, Intrusions and Defenses (RAID)*. DOI: 10.1145/3607199.3607208.

[2] Mohammed Asiri, Neetesh Saxena, Rigel Gjomemo, and Pete Burnap. 2023. Understanding Indicators of Compromise against Cyber-attacks in Industrial Control Systems: A Security Perspective. *ACM Transactions on Cyber-Physical Systems*, 7, 2. DOI: 10.1145/3587255.

[3] Kiran Bandla and Santiago Castro. 2025. APTnotes. https://github.com/aptnotes/data/. (2025).

[4] David Bianco. 2013. The Pyramid of Pain. https://detect-respond.blogspot.com/2013/03/the-pyramid-of-pain.html. (2013).

[5] Xander Bouwman, Harm Griffioen, Jelle Egbers, Christian Doerr, Bram Klievink, and Michel Van Eeten. 2020. A Different Cup of TI? The Added Value of Commercial Threat Intelligence. In *Proceedings of the 29th USENIX Security Symposium*. https://www.usenix.org/conference/usenixsecurity20/presentation/bouwman.

[6] Xander Bouwman, Victor Le Pochat, Pawel Foremski, Tom Van Goethem, Carlos H. Ganan, Giovane C. M. Moura, Samaneh Tajalizadehkhoob, Wouter Joosen, and Michel van Eeten. 2022. Helping Hands: Measuring the Impact of a Large Threat Intelligence Sharing Community. In *Proceedings of the 31st USENIX Security Symposium*. https://www.usenix.org/conference/usenixsecurity22/presentation/bouwman.

[7] Juan Caballero, Gibran Gomez, Srdjan Matic, Gustavo Sánchez, Silvia Sebastián, and Arturo Villacañas. 2023. The Rise of GoodFATR: A Novel Accuracy Comparison Methodology for Indicator Extraction Tools. *Future Generation Computer Systems*, 144. Source code: https://github.com/malicialab/iocsearcher. doi: 10.1016/j.future.2023.02.012.

[8] Ricardo J. G. B. Campello, Davoud Moulavi, and Joerg Sander. 2013. Density-Based Clustering Based on Hierarchical Density Estimates. In *Advances in Knowledge Discovery and Data Mining*. doi: 10.1007/978-3-642-37456-2_14.

[9] Canadian Center for Cyber Security. 2018. Joint Report on Publicly Available Hacking Tools. https://www.cyber.gc.ca/sites/default/files/cyber/publications/joint_report_on_publicly_available_hacking_tools.pdf. (2018).

[10] Hoang Cuong Nguyen, Shahroz Tariq, Mohan Baruwal Chhetri, and Bao Quoc Vo. 2025. Towards Effective Identification of Attack Techniques in Cyber Threat Intelligence Reports using Large Language Models. In *Companion Proceedings of the ACM on Web Conference (WWW)*. doi: 10.1145/3701716.3715469.

[11] EmpireProject. 2025. Empire Post-exploitation Framework. https://github.com/EmpireProject/Empire. (2025).

[12] Facundo Muñoz. 2021. njRAT: A Remote Access Trojan Widely Used by Various Cybercriminals. https://www.welivesecurity.com/la-es/2021/09/29/que-es-njrat-troyano-acceso-remoto-utilizado-cibercriminales/. (2021).

[13] Fraunhofer FKIE. 2025. Silent Chollima. https://malpedia.caad.fkie.fraunhofer.de/actor/silent_chollima. (2025).

[14] FORTRA. 2025. Cobal Strike - Adversary Simulations and Red Team Operations. https://www.cobaltstrike.com/. (2025).

[15] Fraunhofer. 2025. Malpedia. https://malpedia.caad.fkie.fraunhofer.de/. (2025).

[16] Dan Goodin. 2024. Windows Vulnerability Reported by the NSA Exploited to Install Russian Malware. https://arstechnica.com/security/2024/04/kremlin-backed-hackers-exploit-critical-windows-vulnerability-reported-by-the-nsa/. (2024).

[17] Ghaith Husari, Ehab Al-Shaer, Mohiuddin Ahmed, Bill Chu, and Xi Niu. 2017. TTPDrill: Automatic and Accurate Extraction of Threat Actions from Unstructured Text of CTI Sources. In *Proceedings of the 33rd Annual Computer Security Applications Conference (ACSAC)*. doi: 10.1145/3134600.3134646.

[18] Internet Archive. 2025. Wayback Machine. https://web.archive.org/. (2025).

[19] Yuning Jiang, Qiaoran Meng, Feiyang Shang, Nay Oo, Le Thi Hong Minh, Hoon Wei Lim, and Biplab Sikdar. 2025. MITRE ATT&CK Applications in Cybersecurity and The Way Forward. (2025). arXiv: 2502.10825 [cs.CR].

[20] Beomjin Jin, Eunsoo Kim, Hyunwoo Lee, Elisa Bertino, Doowon Kim, and Hyoungshick Kim. 2024. Sharing Cyber Threat Intelligence: Does it Really Help? In *Proceedings of the 31st Network and Distributed System Security Symposium (NDSS)*. doi: 10.14722/ndss.2024.24228.

[21] Robert J Joyce, Dev Amlani, Charles Nicholas, and Edward Raff. 2023. MOTIF: A Malware Reference Dataset with Ground Truth Family Labels. *Computers & Security*, 124. doi: 10.1016/j.cose.2022.102921.

[22] Chaz Lever, Platon Kotzias, Davide Balzarotti, Juan Caballero, and Manos Antonakakis. 2017. A Lustrum of Malware Network Communication: Evolution and Insights. In *Proceedings of the 38th IEEE Symposium on Security and Privacy (S&P)*. San Jose, CA, USA, (May 2017). doi: 10.1109/SP.2017.59.

[23] Pei Han Liao. 2025. Tracking Malware and Attack Expansion: A Hacker Group's Journey across Asia. https://www.fortinet.com/blog/threat-research/tracking-malware-and-attack-expansion-a-hacker-groups-journey-across-asia. (2025).

[24] Martina Lindorfer, Alessandro Di Federico, Federico Maggi, Paolo Milani Comparetti, and Stefano Zanero. 2012. Lines of Malicious Code: Insights Into the Malicious Software Industry. In *Proceedings of the 28th Annual Computer Security Applications Conference (ACSAC)*. Orlando, FL, USA. doi: 10.1145/2420950.2421001.

[25] Federico Maggi, Andrea Bellini, Guido Salvaneschi, and Stefano Zanero. 2011. Finding Non-trivial Malware Naming Inconsistencies. In *Proceedings of the 7th International Conference on Information Systems Security (ICISS)*. doi: 10.1007/978-3-642-25560-1_10.

[26] MalwareLab. 2020. On the Royal Road. https://blog.malwarelab.pl/posts/on_the_royal_road/. (2020).

[27] MISP. 2025. Threat Actors Galaxy. https://github.com/MISP/misp-galaxy/blob/main/clusters/threat-actor.json. (2025).

[28] MITRE. 2025. ATT&CK. https://attack.mitre.org/. (2025).

[29] MITRE. 2024. ATT&CK STIX Data. https://github.com/mitre-attack/attack-stix-data. (2024).

[30] MITRE. 2025. Babuk. https://attack.mitre.org/software/S0638/. (2025).

[31] MITRE. 2025. Conficker. https://attack.mitre.org/software/S0608/. (2025).

[32] MITRE. 2025. SimBad. https://attack.mitre.org/software/S0419/. (2025).

[33] Aziz Mohaisen and Omar Alrawi. 2014. AV-Meter: An Evaluation of Antivirus Scans and Labels. In *Proceedings of the 11th Conference on Detection of Intrusions and Malware & Vulnerability Assessment (DIMVA)*. doi: 10.1007/978-3-319-08509-8_7.

[34] Aleksandr Nahapetyan, Sathvik Prasad, Kevin Childs, Adam Oest, Yeganeh Ladwig, Alexandros Kapravelos, and Brad Reaves. 2024. On SMS Phishing Tactics and Infrastructure. In *Proceedings of the 45th IEEE Symposium on Security and Privacy (S&P)*. doi: 10.1109/SP54263.2024.00169.

[35] NIST. 2025. Official Common Platform Enumeration (CPE) Dictionary. https://nvd.nist.gov/products/cpe. (2025).

[36] OffSec. 2025. Exploit Database. https://www.exploit-db.com/. (2025).

[37] Kris Oosthoek and Christian Doerr. 2019. SoK: ATT&CK Techniques and Trends in Windows Malware. In *Proceedings of the 15th EAI International Conference on Security and Privacy in Communication Systems (SecureComm)*. doi: 10.1007/978-3-030-37228-6_20.

[38] ORKL. 2025. ORKL. https://orkl.eu/. (2025).

[39] Kirsty Paine, Ollie Whitehouse, James Sellwood, and Andrew S. 2023. Indicators of Compromise (IoCs) and Their Role in Attack Defence (RFC 9424). https://datatracker.ietf.org/doc/rfc9424/. (Aug. 2023).

[40] Daniel Plohmann, Martin Clauss, Steffen Enders, and Elmar Padilla. 2017. Malpedia: A Collaborative Effort to Inventorize the Malware Landscape. *The Journal on Cybercrime & Digital Investigations: Proceedings of Botconf*, 3, 1. doi: 10.18464/cybin.v3i1.17.

[41] Md Rayhanur Rahman, Setu Kumar Basak, Rezvan Mahdavi Hezaveh, and Laurie Williams. 2024. Attackers Reveal their Arsenal: An Investigation of Adversarial Techniques in CTI Reports. (2024). arXiv: 2401.01865 [cs.CR].

[42] Md Rayhanur Rahman, Brandon Wroblewski, Mahzabin Tamanna, Imranur Rahman, Andrew Anufryienak, and Laurie Williams. 2024. Towards a Taxonomy of Challenges in Security Control Implementation. In *Proceedings of the 40th Annual Computer Security Applications Conference (ACSAC)*. doi: 10.1109/ACSAC63791.2024.00022.

[43] Nanda Rani, Bikash Saha, Vikas Maurya, and Sandeep Kumar Shukla. 2024. TTPXHunter: Actionable Threat Intelligence Extraction as TTPs from Finished Cyber Threat Reports. *ACM Digital Threats: Research and Practice (DTRAP)*, 5, 4. doi: 10.1145/3696427.

[44] RedHat. 2024. What is a CVE? https://www.redhat.com/en/topics/security/what-is-cve. (Sept. 2024).

[45] Thomas Rid and Ben Buchanan. 2015. Attributing Cyber Attacks. *Journal of Strategic Studies*, 38, 1-2. doi: 10.1080/01402390.2014.977382.

[46] Gonzalo Roa, Manuel Suarez-Roman, and Juan Tapiador. 2025. Hesperus is Phosphorus: Mapping Threat Actor Naming Taxonomies at Scale. (2025). arXiv: 2512.00857 [cs.CR].

[47] Shanto Roy, Emmanouil Panaousis, Cameron Noakes, Aron Laszka, Sakshyam Panda, and George Loukas. 2023. SoK: The MITRE ATT&CK Framework in Research and Practice. (2023). arXiv: 2304.07411 [cs.CR].

[48] Bader Al-Sada, Alireza Sadighian, and Gabriele Oligeri. 2024. MITRE ATT&CK: State of the Art and Way Forward. *ACM Computing Surveys*, 57, 1. doi: 10.1145/3687300.

[49] Aakanksha Saha, Jorge Blasco, Lorenzo Cavallaro, and Martina Lindorfer. 2024. ADAPT it! Automating APT Campaign and Group Attribution by Leveraging and Linking Heterogeneous Files. In *Proceedings of the 27th International Symposium on Research in Attacks, Intrusions and Defenses (RAID)*. doi: 10.1145/3678890.3678909.

[50] Aakanksha Saha, James Mattei, Jorge Blasco, Lorenzo Cavallaro, Daniel Votipka, and Martina Lindorfer. 2025. Expert Insights into Advanced Persistent Threats: Analysis, Attribution, and Challenges. In *Proceedings of the 34th USENIX Security Symposium*. https://www.usenix.org/conference/usenixsecurity25/presentation/saha.

[51] Marcos Sebastián, Richard Rivera, Platon Kotzias, and Juan Caballero. 2016. AVClass: A Tool for Massive Malware Labeling. In *Proceedings of the 19th International Symposium on Research in Attacks, Intrusions and Defenses (RAID)*. doi: 10.1007/978-3-319-45719-2_11.

[52] Sentinel Labs. 2021. ShadowPad: A Masterpiece of Privately Sold Malware RAT. https://assets.sentinelone.com/c/shadowpad?x=p42eqa. (2021).

[53] Sayuj Shah and Vijay K Madisetti. 2025. MAD-CTI: Cyber Threat Intelligence Analysis of the Dark Web Using a Multi-Agent Framework. *IEEE Access*, 13. doi: 10.1109/ACCESS.2025.3547172.

[54] Niraj Shivtarkar and Avinash Kumar. 2022. Lyceum .NET DNS Backdoor. https://www.zscaler.com/blogs/security-research/lyceum-net-dns-backdoor. (2022).

[55] sin5678. 2025. gh0st RAT. https://github.com/sin5678/gh0st. (2025).

[56] Florian Skopik and Timea Pahi. 2020. Under False Flag: Using Technical Artifacts for Cyber Attack Attribution. *Cybersecurity*, 3. doi: 10.1186/s42400-020-00048-4.

[57] Max Smeets. 2016. When Naming Cyber Threat Actors Does More Harm Than Good. https://www.cfr.org/blog/when-naming-cyber-threat-actors-does-more-harm-good. (2016).

[58] Lake E. Strom, Andy Applebaum, Doug P. Miller, Kathryn C. Nickels, Adam G. Pennington, and Cody B Thomas. 2018. MITRE ATT&CK: Design and Philosophy. https://attack.mitre.org/docs/ATTACK_Design_and_Philosophy_March_2020.pdf. (2018).

[59] ThreatMiner. 2025. Data Mining for Threat Intelligence. https://www.threatminer.org/. (2025).

[60] Kevin van Liebergen, Gibran Gomez, Srdjan Matic, and Juan Caballero. 2025. All your (data)base are belong to us: Characterizing Database Ransom(ware) Attacks. In *Proceedings of the 32nd Network and Distributed Systems Security Symposium (NDSS)*. doi: 10.14722/ndss.2025.241887.

[61] Max van der Horst, Ricky Kho, Olga Gadyatskaya, and Michel Mollema. 2025. High Stakes, Low Certainty: Evaluating the Efficacy of High-Level Indicators of Compromise in Ransomware Attribution. In *Proceedings of the 34th USENIX Security Symposium*. https://www.usenix.org/conference/usenixsecurity25/presentation/van-der-horst.

[62] Vector Guo Li and Matthew Dunn and Paul Pearce and Damon McCoy and Geoffrey M. Voelker and Stefan Savage. 2019. Reading the Tea leaves: A Comparative Analysis of Threat Intelligence. In *Proceedings of the 28th USENIX Security Symposium*. https://www.usenix.org/conference/usenixsecurity19/presentation/li.

[63] Apurva Virkud, Muhammad Adil Inam, Andy Riddle, Jason Liu, Gang Wang, and Adam Bates. 2024. How does Endpoint Detection use the MITRE ATT&CK Framework? In *Proceedings of the 33rd USENIX Security Symposium*. https://www.usenix.org/conference/usenixsecurity24/presentation/virkud.

[64] Yiming Wu, Qianjun Liu, Xiaojing Liao, Shouling Ji, Peng Wang, Xiaofeng Wang, Chunming Wu, and Zhao Li. 2021. Price TAG: Towards Semi-Automatically Discovery Tactics, Techniques and Procedures of E-Commerce Cyber Threat Intelligence. *IEEE Transactions on Dependable and Secure Computing*. doi: 10.1109/TDSC.2021.3120415.

# A Implicit Reference Extraction

As discussed in Section 4.2, threat reports may contain implicit references to techniques. We conducted preliminary experiments using large language models (LLMs) to extract techniques from the downloaded threat reports. Specifically, we used the commercial GPT-4 model, guided by a prompt shown in Listing 1, which we experimentally identified as the most effective among other options. To ensure the model analyzed the report content, we removed any tables of techniques included at the end of the report.

Unfortunately, we obtained mixed results as the LLM frequently hallucinated techniques, introducing false positives. An example is ZScaler's 2022 report on the Lyceum group [54]. For this report, `iocsearcher` [7] identified eight techniques (without sub-techniques), all of them in a table at the end of the report. These are the same eight techniques that ATT&CK associates with Lyceum from this report, suggesting that the contributor relied directly on the table for extraction. When we provided the same report after removing the table of identifiers, the LLM returned 11 technique identifiers (seven techniques and four sub-techniques). Of these, only two overlapped with the original table. We manually reviewed the remaining nine, finding that while two were valid, seven were false positives. An example of a correctly extracted implicit reference is *Boot or Logon Autostart Execution: Startup Folder* (T1547.001; part of the *Persistence* tactic) that was extracted from the following text: "written into the Startup folder in order to maintain persistence." An example of a false positive is *Remote System Discovery* (T1018). When asked to justify it, the model responded: "While the report does not specifically mention remote system discovery, many backdoors and malware engage in network reconnaissance, which aligns with T1018."

In future work, we plan to extend the LLM-based extraction and compare it with existing NLP-based approaches [1, 17, 43].

```
I have a detailed threat report written in natural language.
     I need help identifying the TTPs (Tactics, Techniques,
     and Procedures) described in the report and mapping
    them to their corresponding MITRE ATT&CK Technique IDs.
     The output should include:
A list of tactics (high-level strategic goals) based on the
    threat actor's behavior.
A list of techniques (specific actions or behaviors), with
    their corresponding MITRE ATT&CK Technique IDs.
A description of procedures (the exact implementation or
    variation of a technique as described in the report).
For each technique, provide both the name and the MITRE ATT&
    CK Technique ID. Please ensure all identified TTPs are
    clearly mapped to the most relevant MITRE ATT&CK
    entries.
Here's the threat report:
[Insert threat report text here]
The output should be in this JSON format:
Example Output:

{
  "tactics": ["Initial Access", "C2"],
  "techniques": [
    {"name": "Spear Phishing", "MITRE ID": "T1193"},
    {"name": "C2 Channel Over HTTPS", "MITRE ID": "T1071"}
  ],
  "procedures": ["Use of malicious scripts"]
}
```

**Listing 1: Example prompt to instruct an LLM (in our case GPT-4) to identify Tactics, Techniques, and Procedures (TTPs) from a natural language threat report and map them to their respective MITRE ATT&CK technique IDs, formatted as JSON output.**

# B Group Behavior Evolution: Lazarus

As a case study, we investigated the evolution of the Lazarus Group (G0032). Table 10 captures the evolution of the technique profile of Lazarus between 2016 and 2022. Five additional reports published after 2022 refer to Lazarus, but they either analyzed multiple threat groups or did not explicitly mention any technique identifiers.

The profile increases steadily from 40 techniques in 2016 up to 91 in 2022. In 2016, a total of 40 techniques were observed for the group. These include core capabilities such as discovery: *System Network Configuration Discovery* (T1016), *Application Window Discovery (T1010)*, *System Network Configuration* Discovery (T1016), credential access: *Keylogging* (T1056.001), *Password Spraying* (T1110.003), and data destruction: *Disk Content Wipe* (T1561.001).

Between 2017 and 2020, Lazarus introduced a smaller number of new techniques, 2—10 annually. These include techniques for evasion and stealth: *Dynamic-link Library Injection* (T1055.001), *Standard Encoding* (T1132.001), *DLL Side-Loading* (T1574.002) and *Native API* (T1106). In 2021, Lazarus broadened its infrastructure behaviors with techniques *Acquire Infrastructure: Domains* (T1583.001), and *Obtain Capabilities: Digital Certificates* (T1588.004). Finally, in 2022, Lazarus introduced a smaller but technically advanced set of behaviors including in-memory execution to evade detection.

Key techniques include *Dynamic API Resolution* (T1027.007), *Reflective Code Loading* (T1620), which enables obfuscated execution, and *KernelCallbackTable hijacking* (T1574.013), an execution-flow manipulation technique used to hinder analysis.

Out of the total of 91 techniques in the profile, only 6 are considered as group-specific in ATT&CK v15.1 (released in April 2024), the first one being *Disk Wipe: Disk Content Wipe* (T1561.001) reported in 2016. In 2022, five new supposedly Lazarus-specific defense evasion behaviors were reported: *Obfuscated Files or Information: Dynamic API Resolution* (T1027.007), *Indirect Command Execution* (T1202), *System Binary Proxy Execution* (T1218), *Hijack Execution Flow: KernelCallbackTable* (T1574.013), and *Reflective Code Loading* (T1620). However, if we consider ATT&CK v18.1 (released in October 2025), five of the supposedly group-specific Lazarus techniques have been reported as being used by other groups and only *Hijack Execution Flow: KernelCallbackTable* (T1574.013) remains specific to Lazarus.

**Table 10: Evolution of the Lazarus technique profile over time based on ATT&CK v15.1. For each year, we list the number of reports, techniques in those reports (cumulative in parenthesis), and group-specific techniques (cumulative in parenthesis). *In the latest ATT&CK v18.1, only one technique remains group-specific.**

| Year | Reports | Techniques | Group-Specific Techniques | |
|---|---|---|---|---|
| 2016 | 5 | +40 (40) | +1 (1) | T1561.001 |
| 2017 | 2 | +2 (42) | 0 (1) | – |
| 2018 | 4 | +28 (52) | 0 (1) | – |
| 2019 | 1 | +6 (56) | 0 (1) | – |
| 2020 | 4 | +9 (58) | 0 (1) | – |
| 2021 | 3 | +26 (78) | 0 (1) | – |
| 2022 | 2 | +30 (91) | +5 (6) | T1620, T1202, T1218, T1027.007, T1574.013* |